\documentclass[showpacs,preprintnumbers, amsmath,amssymb,superscriptaddress,floatfix]{revtex4}
\usepackage{graphicx}
\begin{document}
\newcommand {\bb}{\bibitem}
\newcommand {\be}{\begin{equation}}
\newcommand {\ee}{\end{equation}}
\newcommand {\bea}{\begin{eqnarray}}
\newcommand {\eea}{\end{eqnarray}}
\newcommand {\nn}{\nonumber}

\title{Topological defects in triplet superconductors UPt$_{3}$,
Sr$_{2}$RuO$_{4}$, etc.}

\author{K. Maki}
\author{S. Haas}
\author{D. Parker}

\address{Department of Physics and Astronomy, University of Southern
California, Los Angeles, CA 90089-0484 USA}

\author{H. Won}

\address{Department of Physics, Hallym University,
Chuncheon 200-702, South Korea}

\begin{abstract}
After a brief introduction on nodal superconductors, we review the topological
defects in triplet superconductors such as UPt$_{3}$, Sr$_{2}$RuO$_{4}$, etc.
This is in part motivated by the surprising discovery of Ana Celia Mota and
her colleagues that in some triplet superconductors the flux motion is
completely impeded (the ideal pinning).
Among topological defects the most prominent is Abrikosov's vortex with
quantum flux $\phi_{0}= \frac{hc}{2e}$.  Abrikosov's vortex is universal
and ubiquitous and seen in both conventional and unconventional
superconductors by the Bitter decoration technique, small angle
neutron scattering (SANS), scanning tunneling microscopy (STM), 
micromagnetometer and more recently by Lorentz electron micrograph.
In order to interpret the experiment by Mota et al a variety of textures
are proposed.  In particular, in analogy to superfluid $^{3}$He-A the 
$\hat{\ell}$-soliton and $\hat{d}$-soliton play the prominent role.  We
review these notions and point out possible detection of these domain
walls and half-quantum vortices in some triplet superconductors.

\end{abstract}
\maketitle
\section{Introduction}

We shall first survey the new world of nodal superconductors, which 
appeared on the scene in 1979.  Indeed nodal superconductors are a child
of the 21st century \cite{1,2,3}.  Although the presence of nodal 
superconductors in heavy-fermion superconductors like CeCu$_{2}$Si$_{2}$,
UPt$_{3}$, UBe$_{13}$ and others was found in the late eighties \cite{1},
the systematic study of the gap function $\Delta({\bf k})$ began only
after the discovery of the high-T$_{c}$ cuprates La$_{2-x}$Ba$_{x}$CuO$_{4}$
by Bednorz and M\"{u}ller \cite{4} in 1986.  The d-wave symmetry of
high-T$_{c}$ cuprates YBCO, Bi-2212, etc. was established circa 1994 through
the elegant Josephson interferometry \cite{5,6} and the powerful angle resolved
photoemission spectrum (ARPES) \cite{7} among others.  In 1993 Volovik
\cite{8} derived the quasiparticle density of states of the vortex state
in nodal superconductors within the semiclassical approximation.  The
surprising $\sqrt{H}$ dependence of the specific heat has been seen in
YBCO \cite{9}, LSCO \cite{10}, and Sr$_{2}$RuO$_{4}$\cite{11,12}.  
Later Volovik's
approach was extended in a variety of directions: a) the study of the 
thermal conductivity \cite{13,14}; b) for an arbitrary field direction
\cite{15}; and c) for different classes of $\Delta({\bf k})$ \cite{16}.
These are summarized in Ref.[17].  Until now the powerful ARPES and
Josephson interferometry have not been applied outside of high-T$_{c}$ cuprate
superconductors.

Since 2001 Izawa et al
have determined the gap function $\Delta({\bf k})$ in Sr$_{2}$RuO$_{4}$\cite{18},CeCoIn$_{5}$ \cite{19}, $\kappa-$(ET)$_{2}$Cu(NCS)$_{2}$ \cite{20},
YNi$_{2}$B$_{2}$C \cite{21}, and PrOs$_{4}$Sb$_{12}$ \cite{22,23}
via the angle-dependent magnetothermal conductivity.  
These $|\Delta({\bf k})|$'s are shown in Figure 1.
\begin{figure}
\includegraphics[width=2cm]{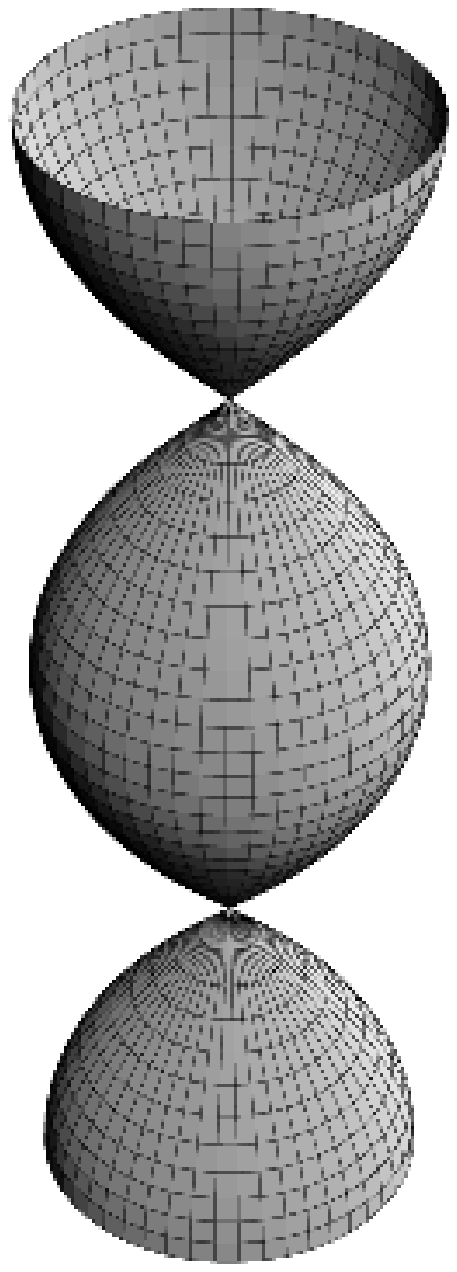}
\includegraphics[width=4cm]{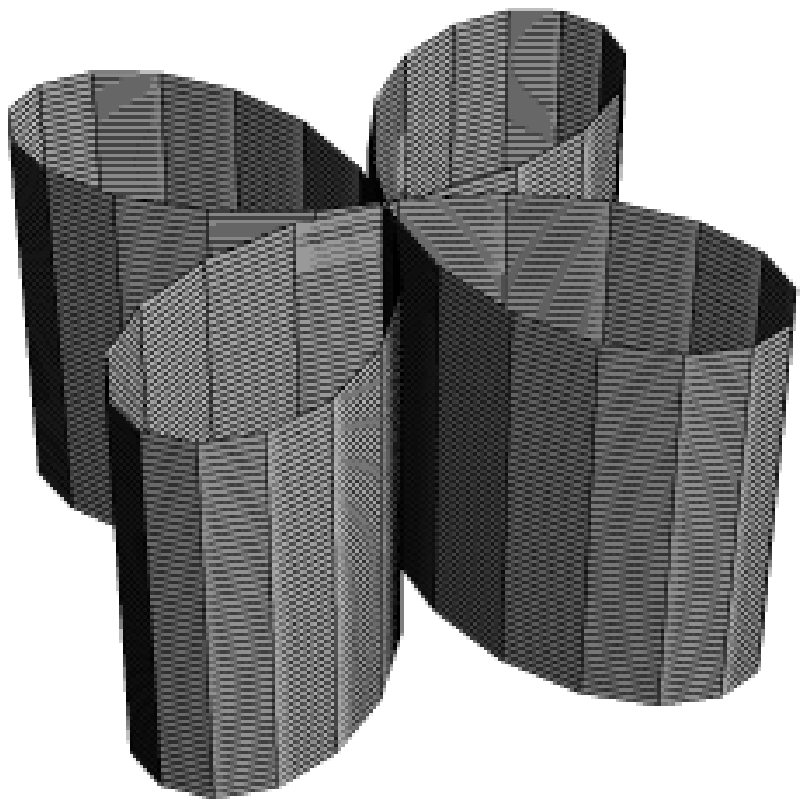}
\includegraphics[width=4cm]{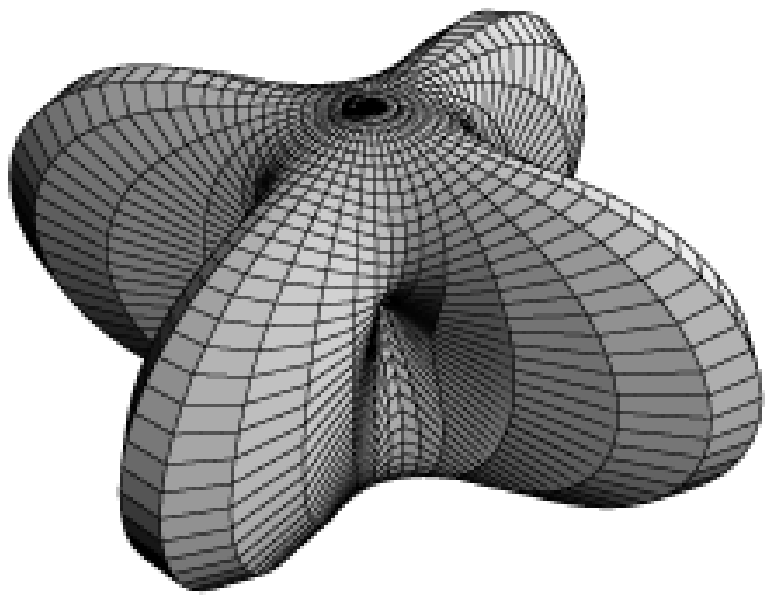}
\includegraphics[width=3.2cm]{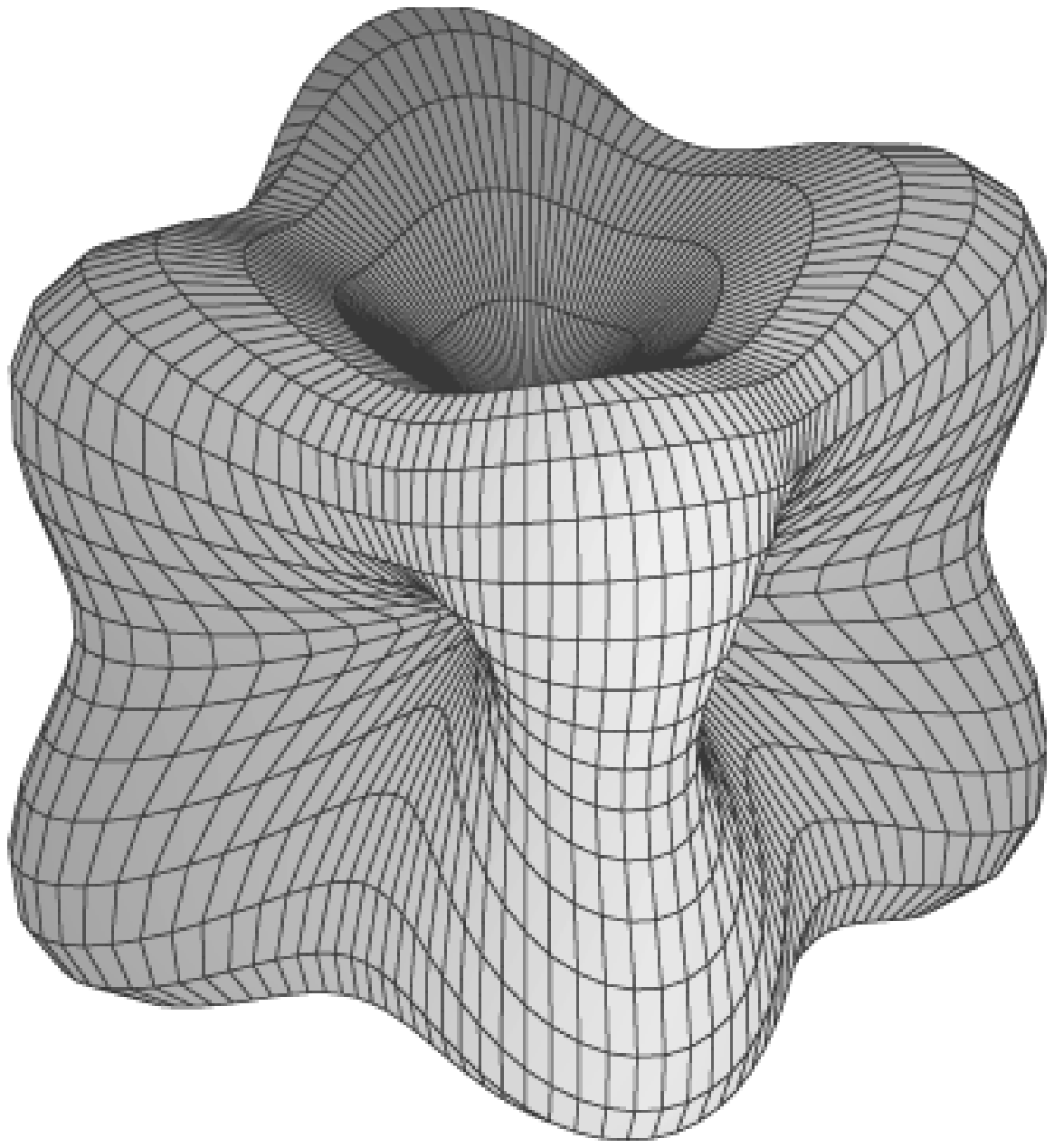}
\includegraphics[width=4cm]{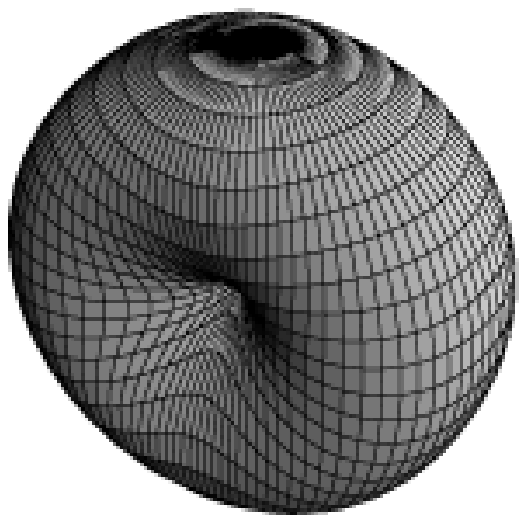}
\caption{From top left, order parameters for Sr$_{2}$RuO$_{4}$, CeCoIn$_{5}$ and $\kappa-$(ET)$_{2}$Cu(NCS)$_{2}$, YNi$_{2}$B$_{2}$C, PrOs$_{4}$Sb$_{12}$ (A and B phases)}
\end{figure}
In addition, the gap function of UPt$_{3}$ was established around 1994-6
as E$_{2u}$ through the anisotropy in the thermal conductivity
\cite{24} and the constancy of the Knight shift in NMR \cite{25}.  Somewhat
surprisingly all these superconductors are nodal and their quasiparticle
density of states increases linearly in $|E|$ for $|E|/\Delta \ll 1$:
\bea
N(E) & \sim & |E|/\Delta
\eea
For example, this implies that the p-wave superconductivity in Sr$_{2}$RuO$_{4}$
as proposed in Ref. 26 is not consistent with the specific heat data
\cite{11}.  Also as discussed elsewhere \cite{27,28}, the two gap model is
of little help in this matter.

More recently the quasiparticle density of states in the vortex state in
Sr$_{2}$RuO$_{4}$ has been reported \cite{29}.  Indeed the observed
quasiparticle density of states is very consistent with that predicted 
for an f-wave order parameter \cite{30}.  Also many of these superconductors
are triplet:  UPt$_{3}$, 
Sr$_{2}$RuO$_{4}$, (TMTSF)$_{2}$PF$_{6}$,  U$_{1-x}$Th$_{x}$Be$_{13}$,
URu$_{2}$Si$_{2}$, PrOs$_{4}$Sb$_{12}$, UNi$_{2}$Al$_{3}$ and CePt$_{3}$Si,
for example. 

\section{Textures in triplet superconductors}

Here we consider possible textures in triplet superconductors.  For 
simplicity we concentrate on two f-wave superconductors: UPt$_{3}$
(3-dimensional) and Sr$_{2}$RuO$_{4}$ (quasi 2-dimensional), which
we understand well.  Mota et al \cite{33,34} have discovered the
ideal pinning in the B phase of UPt$_{3}$ and below T=70 mK in 
Sr$_{2}$RuO$_{4}$.  Also these systems are characterized by $\hat{\ell}$ and
$\hat{d}$ similarly to superfluid $^{3}$He-A \cite{35,36}.  Here $\hat{\ell}$
is the direction of the pair angular momentum.  But unlike superfluid
$^{3}$He-A $\hat{\ell}$ is fixed to be parallel to one of the crystal
axes: $\hat{\ell} \parallel \pm \hat{c}$.  $\hat{\ell}$ can be called
the chiral vector.  On the other hand $\hat{d}$ describes the spin
configuration of the pair and is perpendicular to the pair spin 
${\bf S_{pair}}$.  In the equilibrium configuration $\hat{d} \parallel
\pm \hat{\ell}$ in superfluid $^{3}$He-A, 
UPt$_{3}$ and Sr$_{2}$RuO$_{4}$
\cite{26}.  Also we believe that some triplet superconductors may not have
$\hat{\ell}$ and $\hat{d}$ vectors, as observed in superfluid $^{3}$He-B.  So it
is very important to know if the superconductivity breaks the chiral
symmetry, as observed in the experiments by Mota et al \cite{33,34} on the
B phase in U$_{1-x}$Th$_{x}$Be$_{13}$ (see Fig. 2).
\begin{figure}
\includegraphics[width=9cm]{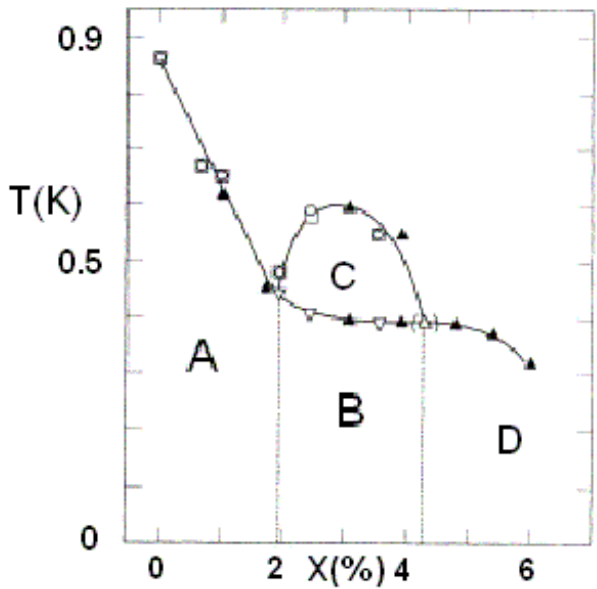}
\caption{Phase diagram of U$_{1-x}$Th$_{x}$Be$_{13}$ 
from Heffner et al \cite{heffner}.}
\end{figure}
But it is possible that the superconductor in UBe$_{13}$ does not have
$\hat{\ell}$ and $\hat{d}$.  Also we conjecture that the superconductivity
in the B phase of PrOs$_{4}$Sb$_{12}$ has $\hat{\ell}$ and $\hat{d}$
vectors \cite{23} and that probably all other quasi 2-dimensional systems
such as URu$_{2}$Si$_{2}$, UNi$_{2}$Al$_{3}$ and CePt$_{3}$Si do as well.
Therefore we can think of a variety of domain walls as in superfluid $^{3}$
He-A \cite{38,39}.  Indeed the $\hat{\ell}$-soliton in Sr$_{2}$RuO$_{4}$ 
has been considered by Sigrist and Agterberg \cite{40}. 
The $\hat{\ell}$-soliton is created when in one side of the wall $\hat{\ell}
\parallel \hat{c}$, while in the other side $\hat{\ell} \parallel -\hat{c}$.
The chirality changes across the $\hat{\ell}$-soliton.  However, unlike
in superfluid $^{3}$He-A $\hat{\ell}$ is practically fixed to be parallel
to $\pm \hat{c}$.  For example, when $\hat{\ell}$ is in the ab-plane
there will be little superconducting order parameter left.  Therefore we can
estimate the $\hat{\ell}$-soliton energy per unit area as 
\bea
f_{\hat{\ell}} & \simeq & \frac{1}{4}N_{0}\Delta^{2}(T)\xi(T) \\
& \simeq & \frac{1}{4}N_{0}v_{F}\Delta(T)
\eea
where $\Delta(T)$ is the maximum value of the energy gap and $v_{F}$
is the Fermi velocity.

On the other hand the $\hat{d}$-soliton may be much more easily created
\cite{41}.  The crucial element here is the spin-orbit energy which binds
$\hat{d}$ parallel to $\hat{\ell}$.  The relevant energy can 
deduced from the NMR data of UPt$_{3}$ \cite{42} and Sr$_{2}$RuO$_{4}$ 
\cite{43}.  We estimate
$\Omega(T)/\Delta(T)$ for UPt$_{3}$ and Sr$_{2}$RuO$_{4}$: 
$\Omega(T)/\Delta(T) \sim 0.5 \times 10^{-3}$ (B phase) and
$0.2 \times 10^{-4}$ respectively.  Here $\Omega(T)$ is the characteristic
frequency associated with the $\hat{\ell}$ and $\hat{d}$ coupling \cite{41}.
Then the areal energy for the $\hat{d}$-soliton is given by
\bea
f_{\hat{d}} & \simeq & \frac{1}{4}N_{0}v_{F}\Omega(T).
\eea
This is smaller than $f_{\hat{\ell}}$ for UPt$_{3}$ and Sr$_{2}$RuO$_{4}$
by factors of $0.5 \times 10^{-3}$ and $0.2 \times 10^{-4}$, respectively.
Also $\xi_{d}= v_{F}/\Omega(T)$ gives the spatial extension of the
domain well.  This ranges from $10 \mu$m $\sim$ 1 mm.

In the presence of $\hat{\ell}$-solitons vortices may enter into the
superconducting state as observed in the vortex sheet of superfluid $^{3}$He-A 
\cite{44,45}.  Otherwise the motion of the vortex across the $\hat{\ell}$-
soliton is completely impeded as discussed in \cite{40}. When an 
Abrikosov vortex encounters a $\hat{d}$-soliton, the vortex appears to
split into two half-quantum vortices (HQV) in the vicinity of $T \simeq T_{c}$.
At lower temperatures it appears that Abrikosov's vortex should tunnel through
the $\hat{d}$-soliton.  This type of HQV was first predicted in the context
of superfluid $^{3}$He-A \cite{46,47,48}.  However, these HQV's have not
yet been observed in superfluid $^{3}$He-A \cite{49}.  Therefore 
half-quantum vortices may be first observed in triplet superconductors.

\section{Half-quantum vortices}

Here we shall consider a pair of half-quantum vortices (HQV) attached
to a $\hat{d}$-soliton \cite{41}.  The texture free energy is given by
\bea
f_{pair} & = & \frac{1}{2}\chi_{N}C^{2}\int dy\,dz(K(\nabla\phi)^{2} +
\sum_{i,j} |\partial_{i}\hat{d_{j}}|^{2}+\xi_{d}^{-2}\sin^{2}(\Psi))
\eea
where C is the spin wave velocity, $\xi_{d}=C(T)/\Omega(T)$ and
\bea
K & = & \frac{\rho_{s}}{\rho_{s,spin}} = \frac{1+\frac{1}{3}F_{1}}
{1+\frac{1}{3}F_{1}^{a}} \frac{1+\frac{1}{3}F_{1}^{a}(1-\rho_{s}^{0})}
{1+\frac{1}{3}F_{1}(1-\rho_{s}^{0})}
\eea
where $\rho_{s}$ and $\rho_{s,spin}$ are the superfluid density and the
spin superfluid density respectively.  Here $F_{1}$ and $F_{1}^{a}$ are
the Landau parameters and $\rho_{s}^{0}(T)$ is the superfluid density
when $F_{1}=F_{1}^{a}=0$.  The superfluid density for UPt$_{3}$ and
Sr$_{2}$RuO$_{4}$ are shown in Fig. 3.  In this analysis we assumed
\begin{figure}
\includegraphics[width=9cm]{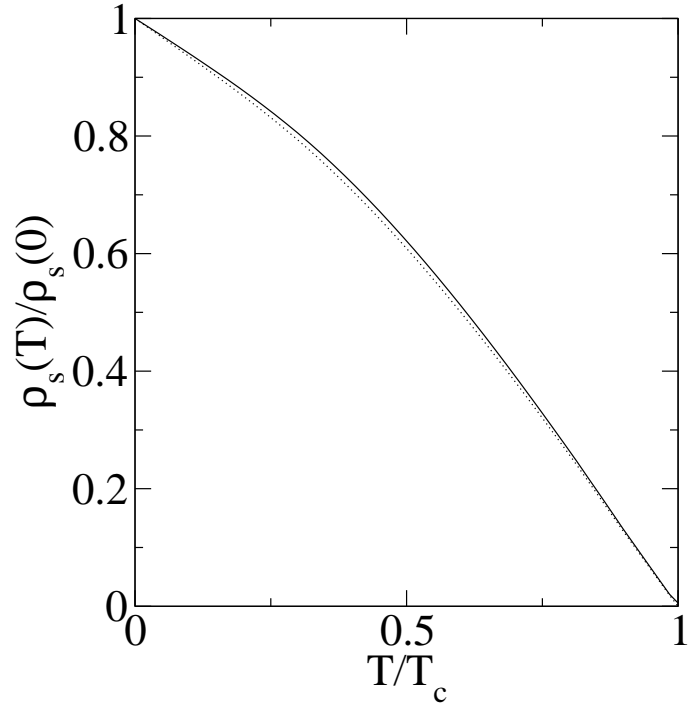}
\caption{The superfluid density for the B phase of UPt$_{3}$ (solid line) and 
Sr$_{2}$RuO$_{4}$ are shown}
\end{figure}
that $|\Delta({\bf k})| \sim |\cos\theta \sin^{2}\theta|$ and 
$\sim |\cos \chi|$ in UPt$_{3}$ \cite{53} and Sr$_{2}$RuO$_{4}$ respectively.
We note that $\rho_{s}^{0}$ in Sr$_{2}$RuO$_{4}$ is the same as in d-wave
superconductivity \cite{54}.  In particular, for T in the vicinity of T$_{c}$
we obtain
\bea
\rho_{s}^{0} & \simeq & -\frac{4}{3}\ln(\frac{T}{T_{c}}) 
\mathrm{\,\,for \,\, Sr_{2}RuO_{4}} \\
& \simeq & - \frac{143}{105}\ln(\frac{T}{T_{c}}) \mathrm{\,\,for \,\, UPt_{3}}
\eea
respectively.  Here we assumed that ${\bf H} \parallel \hat{a}$ and that 
the domain
wall extends in the y-z plane.  Here we consider 2 typical cases as shown
in Fig. 4a) and b).  In Fig. 4a) the $\hat{d}$-soliton runs parallel
to the c axis while in Fig. 4b) it runs parallel to the b-axis.
\begin{figure}
\includegraphics[width=4.5cm]{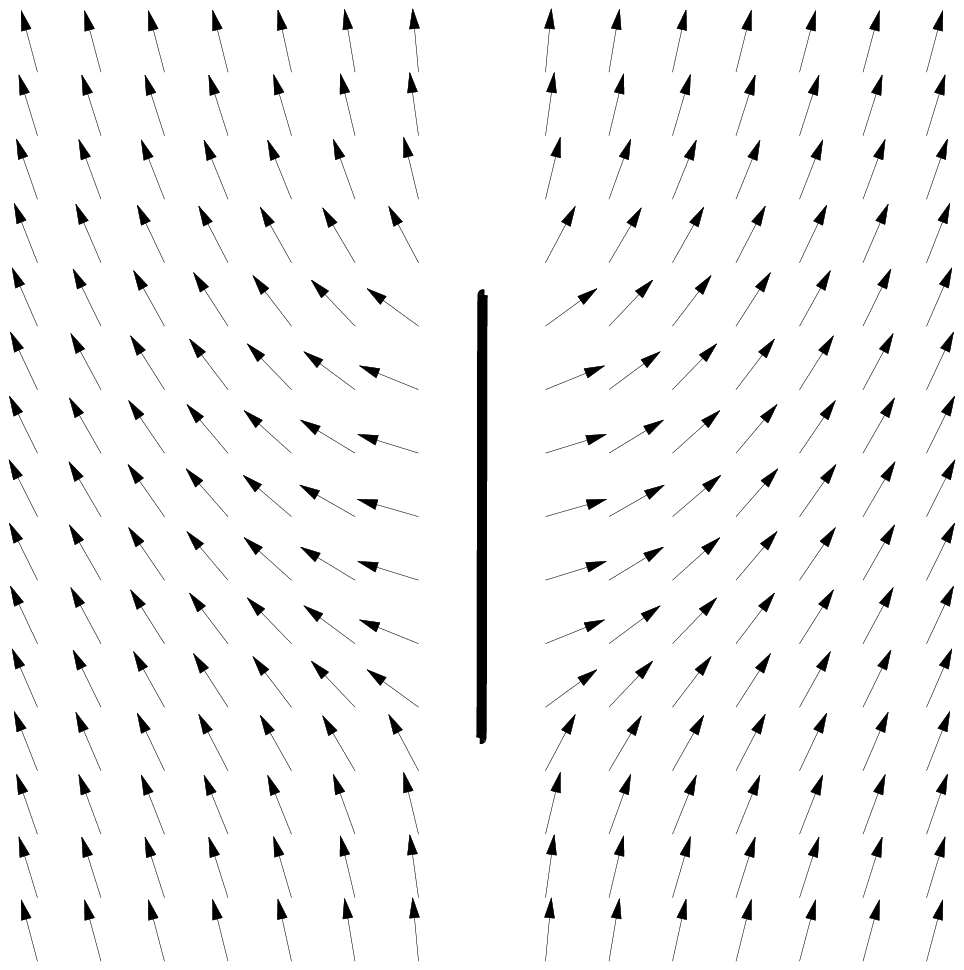}
\includegraphics[width=4.5cm]{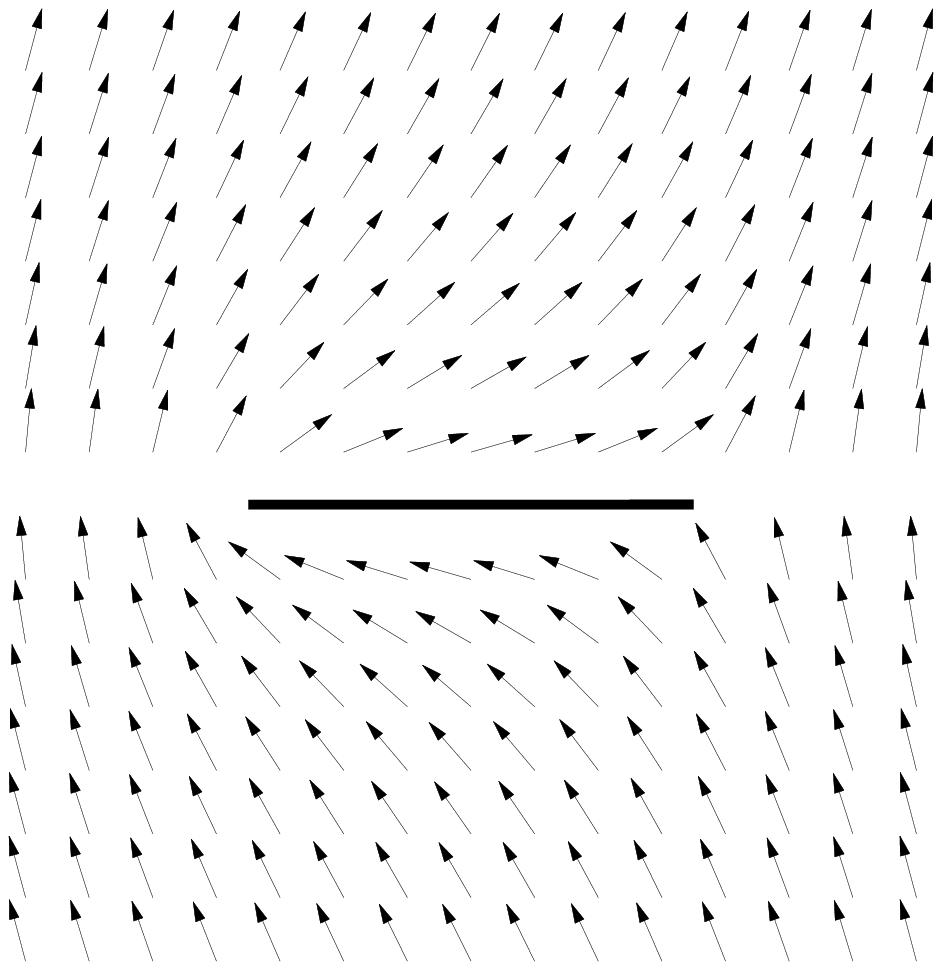}
\caption{The spatial orientation of the $\hat{d}$-vector}
\end{figure}
For the first configuration (i.e. a) we parametrize 
$\hat{d}= \cos \psi \hat{z} +
\sin \psi \hat{y}$ with
\bea
\psi(y,z) &=& \frac{1}{2}(\arctan(\frac{z+(R/2)}{y}) - 
\arctan(\frac{z-(R/2)}{y}))
\eea
with 2 HQV located at $(y,z) = (0,R/2)$ and $(0,-R/2)$ while in the second
configuration (b)
\bea
\psi(y,z) &=& \frac{1}{2}(\arctan(\frac{y+(R/2)}{z}) + 
\arctan(\frac{y-(R/2)}{z}))
\eea
Also $\Phi$ in Eq.(7) is the phase of the order parameter $\Delta({\bf k})$.
Then the total free energy reduces to 
\bea
f_{pair}&=& \frac{1}{2}\chi_{N}C^{2}(\pi K \ln(\lambda/R) + \pi \ln(R/\xi)
+\pi (\frac{R}{2\xi_{d}})^{2} \times \ln (4\xi d/R)
\eea
where $\lambda$ and $\xi$ are the magnetic penetration depth 
and the coherence length respectively.

By minimizing $f_{pair}$ with respect to R, we obtain
\bea
R_{0}^{2} &=& 2\xi_{d}^{2}(K-1)/[\ln(\frac{4\xi f}{\sqrt{e}R_{0}})] > 0
\eea
where $R_{0}$ is the optimal distance of a pair of HQVs.  First in order
to have a pair of HQVs we need $K>1$, which is guaranteed 
when $F_{1}>F_{1}^{a}$ and $T \simeq T_{c}$.  Also it is necessary to have
\bea
R_{0} < \frac{4}{\sqrt{e}}\xi_{d}
\eea
In particular when $K-1 \ll 1$, we obtain
\bea
R_{0}/\xi_{d} & \simeq & \frac{4}{\sqrt{e}} - \frac{\sqrt{e}}{2}(K-1)
\eea
where $e= 2.71828\ldots$.  Also the separation betweeen a pair of
HQV should be of the order of $\xi_{d} \sim 10 \mu$m $\sim$ 1 mm.

\section{Concluding Remarks}

We have described an abundance of triplet superconductors.  Many of
their order parameters possess the $\hat{\ell}$ and $\hat{d}$ vectors:
$\Delta({\bf k})$ in the B phase of UPt$_{3}$, Sr$_{2}$RuO$_{4}$, both
the A and B phase of PrOs$_{4}$Sb$_{12}$ and perhaps many other systems.
In these systems the $\hat{d}$-soliton is the most common domain wall.  The
presence of the $\hat{d}$-soliton can impede the flux motion in a variety
of ways.  The most intriguing is the splitting of 
an Abrikosov vortex into a pair
of half-quantum vortices as discussed in \cite{41}.  We expect that some of the
techniques used to observe Abrikosov's vortex can be used in the present
circumstances.  These techniques include the Bitter decoration technique 
\cite{55}, small angle neutron scattering (SANS) \cite{56}, scanning tunneling
microscopy (STM)\cite{57}, micromagnetometer \cite{58} 
and more recently Lorentz
electron micrograph \cite{59}.  We expect the exploration of these topological
defects in triplet superconductors will enhance our understanding of these
exotic superconductors.  Also they will provide ideal laboratories to check
rich field-theoretical concepts at moderately low temperatures from $1 
\sim 100 \,$ mK.  Therefore the future of topological defects in nodal
superconductors is still wide open.

\section*{Acknowledgments}

We have benefitted from collaborations with Balazs Dora, Hae-Young Kee,
Yong Baek Kim, Andr\'as V\'anyolos and Attila Virosztek on related
subjects.

\end{document}